\documentclass[aps,prl,twocolumn,superscriptaddress,showpacs,floatfix]{revtex4-1}

\usepackage{amsmath,amssymb,amsfonts,float,graphics,epsfig,epstopdf,color,verbatim,tabularx,bm,multirow,appendix}
\usepackage[utf8]{inputenc}
\usepackage[T1]{fontenc}
\usepackage{xcolor}

\usepackage{dsfont}
\usepackage{textcomp}
\usepackage{yfonts}
\usepackage{bm}
\usepackage{subfigure}
\usepackage{mathrsfs}
\usepackage{graphicx}
\usepackage{verbatim}
\usepackage{hyperref}
\usepackage{braket}
\usepackage[normalem]{ulem}
\usepackage{tikz}
	\usetikzlibrary{calc}
	\usetikzlibrary{shapes.multipart}

\newcommand{\comments}[1]{}

\makeatletter
\def\l@subsubsection#1#2{}
\makeatother

\begin{document}

\title{Bipartite entanglement and surface criticality: \\The extra contribution of non-ordinary edge in entanglement}

\author{Yanzhang Zhu}
\affiliation{State Key Laboratory of Surface Physics and Department of Physics, Fudan University, Shanghai 200433, China}
\affiliation{Department of Physics, School of Science and Research Center for Industries of the Future, Westlake University, Hangzhou 310030, China}
\affiliation{Institute of Natural Sciences, Westlake Institute for Advanced Study, Hangzhou 310024, China}

\author{Zenan Liu}
\affiliation{Department of Physics, School of Science and Research Center for Industries of the Future, Westlake University, Hangzhou 310030, China}
\affiliation{Institute of Natural Sciences, Westlake Institute for Advanced Study, Hangzhou 310024, China}

\author{Zhe Wang}
\email{wangzhe90@westlake.edu.cn}
\affiliation{Hangzhou International Innovation Institute, Beihang University, Hangzhou 311115, China}
\affiliation{Department of Physics, School of Science and Research Center for Industries of the Future, Westlake University, Hangzhou 310030, China}
\affiliation{Institute of Natural Sciences, Westlake Institute for Advanced Study, Hangzhou 310024, China}

\author{Yan-Cheng Wang}
\email{ycwangphys@buaa.edu.cn}
\affiliation{Hangzhou International Innovation Institute, Beihang University, Hangzhou 311115, China}
\affiliation{Tianmushan Laboratory, Hangzhou 311115, China}

\author{Zheng Yan}
\email{zhengyan@westlake.edu.cn}
\affiliation{Department of Physics, School of Science and Research Center for Industries of the Future, Westlake University, Hangzhou 310030, China}
\affiliation{Institute of Natural Sciences, Westlake Institute for Advanced Study, Hangzhou 310024, China}

\begin{abstract}
Recent works on the scaling behaviors of entanglement entropy at the $SO(5)$ deconfined quantum critical point (DQCP) sparked a huge controversy. Different bipartitions gave out totally different conclusions for whether the DQCP is consistent with a unitary conformal field theory. In this work, we connect two previously disconnected fields -- the many-body entanglement and the surface criticality -- to reveal the behaviors of entanglement entropy in various bipartite scenarios, and point out that only the ordinary bipartition purely reflects the criticality of the bulk; otherwise, the extra gapless edge mode will also contribute to the entanglement. We have demonstrated that the correspondence between the entanglement spectrum and the edge energy spectrum still approximately persists even at a bulk-gapless point, thereby influencing the behavior of entanglement entropy. Our results establish that boundary conditions induced by the cut are decisive for entanglement-based probes and provide practical protocols to separate bulk from boundary contributions.
\end{abstract}
\date{\today}
\maketitle

\textit{\color{blue} Introduction.---}
Beyond the Landau-Ginzburg-Wilson paradigm, a deconfined quantum critical point (DQCP) describes a continuous phase transition between two unrelated symmetry-breaking phases. Such transitions display unusual criticality and potentially host emergent higher symmetries~\cite{senthil2004deconfined,sandvik2007evidence,sandvik2010continuous,Nahum2015PRL,ma2018dynamical,shao2016quantum,wang2017deconfined,senthil2024deconfined,mao2025detecting,yang2025deconfined,liu2025edge}. Several models have become the standard testbeds for deconfined‐criticality scenarios—including the nonlinear sigma models with topological term~\cite{ma2020theory,lu2023nonlinear,wang2021phases}, the $J$-$Q$ models~\cite{sandvik2007evidence,d2024entanglement,deng2024diagnosing,lou2009antiferromagnetic} and some other boson or fermion models~\cite{Liu_2019,liaoGross2022,zhang2018continuous,liu2024deconfined} etc. However, whether the DQCP is actually a weakly first-order phase transition or real critical point is a long-standing problem.

R\'enyi entanglement entropy (EE) provides a powerful lens on such quantum phase transitions, as its scaling can reveal underlying conformal field theory (CFT) characteristics~\cite{xu2011entanglement,chen2015scaling,elvang2015exact,wang2025probing}. In $(2+1)$ D systems, CFT predicts that EE at a critical point follows an area law ($s=al+b\ln l+c$) with universal logarithmic corrections originating from corners of the entanglement boundary, and the constant corrections also usually carry universal information~\cite{fradkin2006entanglement,casini2007universal,metlitski2009entanglement,laflorencie2016quantum}. Indeed, identifying these corner contributions and constant corrections in numerical simulations is a promising route to establishing the field-theoretic description of a given transition~\cite{zhao2022measuring,wang2025bipartite,ding2025tracking}. 

However, extracting the universal EE behavior from a potential DQCP has proven to be subtle. A quantum Monte Carlo (QMC) study on the $J$-$Q_3$ model with emergent $SO(5)$ DQCP found a coefficient of the cornered correction term that violates the prediction of unitary CFT, and interpreted this as evidence for a weakly first-order phase transition~\cite{zhao2022scaling}. The same diagnostic subsequently distinguished weakly first-order versus genuine DQCP behavior across $SU(N)$ generalizations~\cite{song2025evolution}. More recent studies uncovered an additional logarithmic term even for a cornerless bipartition~\cite{song2024extracting,deng2024diagnosing}. That logarithmic term has now been attributed to Goldstone modes of an emergent $SO(5)$ symmetry breaking, once finite-size effects are handled carefully~\cite{deng2024diagnosing,deng2023improved}.

\begin{figure}
    \centering
    \includegraphics[width=0.9\linewidth]{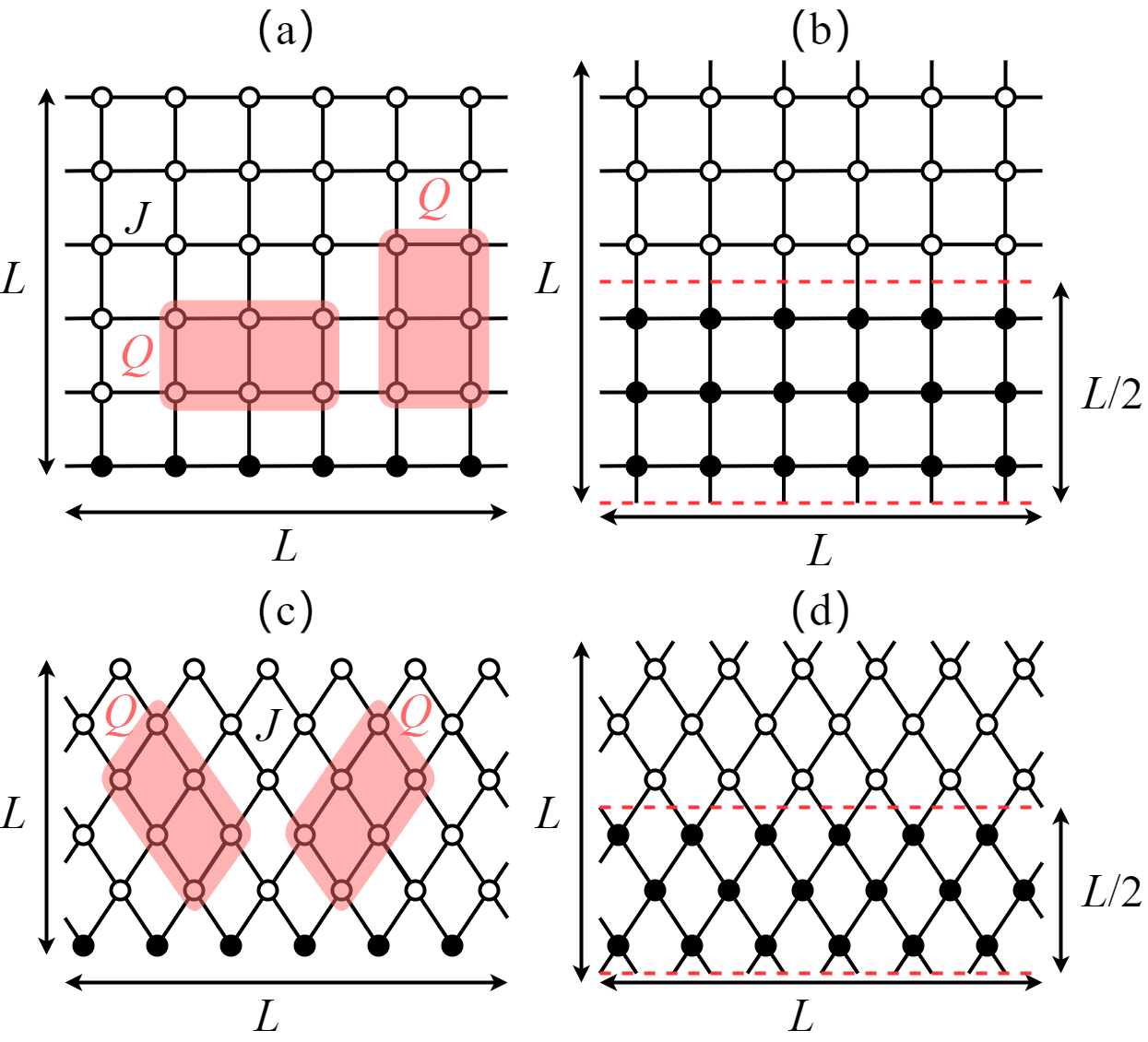}
    \caption{Lattices of $J$-$Q_3$ model with standard/tilted cut considered in this work. The system sizes are set the same vertically and horizontally. The black dots refers to the spins where the edge excitation/entanglement spectra are derived. (a)(b) $J$-$Q_3$ model with standard cut/bipartition. (c)(d) $45^\circ$ tilted cut/bipartition. The boundaries in (a)(c) are vertically open and horizontally periodic. (b)(d) are periodic in both directions, with the red dashed lines referring to the entanglement subsystem edges.}
    \label{fig:model}
\end{figure}

A sharper controversy arose when Ref.~\cite{d2024entanglement} showed that an unusually oriented bipartition, a $45^{\circ}$ tilted cut (as shown in Fig.~\ref{fig:model}(c)) produces EE scaling in striking agreement with an $SO(5)$ criticality, whereas the standard (axis-aligned) cut (shown in Fig.~\ref{fig:model}(a)) does not. The authors argued that only the tilted cut treats all four columnar valence-bond solid (VBS) configurations equally to guarantee emergent symmetry. It shows that the discrepancy of the entanglement entropies is purely a geometrical artifact of bipartitions.

At a fixed bulk critical point, CFT predicts a universal bulk contribution to EE, which seemingly does not depend on the cut. However, in the field of surface criticality, different physical boundaries can introduce additional edge modes (relevant boundary operators) and lead to different edge criticalities \cite{ding2018,weber2018nonordinary,weber2019,wang2022bulk,wang2023extraordinary,zhu2021surface,wang2024unconventional,Jian2021,ding2023special}, which inspires us to explore whether there are similar phenomena in an entanglement cut.

Through large-scale stochastic series expansion (SSE) QMC simulations of the $J$-$Q_3$ model, we show that the tilted bipartition introduces additional physics at the physical/entanglement boundary, which mimics CFT-like bulk behavior but in fact arises from an edge effect. We compare the low-lying entanglement spectrum (ES) from a ``fake'' entanglement cut to the actual energy spectrum of a physical edge (a real cut), thereby directly demonstrating that the Li-Haldane-Poilblanc conjecture (ES resembles edge energy spectrum) is still approximately valid even when the bulk is gapless~\cite{li2008entanglement,poilblanc2010entanglement,liu2024demonstrating,li2024relevant,yu2024universal}. 
In other words, only the EE with an ordinary cut purely reflects the bulk criticality. Moreover, we have discovered a new set of critical exponents corresponding to the bulk criticality of the DQCP, which satisfy the existing scaling relations. We note that while a first-order phase transition may occur in the thermodynamic limit of the $J$-$Q_3$ model we simulated here, finite-size systems still capture the properties of the nearby DQCP.

\textit{\color{blue} Model and methods.---}
We focus on the spin-1/2 $J$-$Q_3$ model on the square lattice~\cite{sandvik2007evidence,lou2009antiferromagnetic}. The Hamiltonian reads,
\begin{equation}
	H=-J\sum_{\langle ij \rangle}P_{i,j}-Q\sum_{\langle ijklmn \rangle}P_{ij}P_{kl}P_{mn},
\end{equation}
where $P_{ij}=\frac{1}{4}-\mathbf{S}_{i}\cdot\mathbf{S}_{j}$ is the two-spin singlet projector. $J > 0$ is the nearest-neighbor antiferromagnetic (AFM) Heisenberg coupling, and $Q>0$ is the six-spin interaction, as illustrated in Fig.~\ref{fig:model}.  Setting $J=1$, the ground state exhibits the AFM order for small $Q$ and the VBS order for large $Q$.  At the phase transition points $Q_c=1.49153$ the system emerges a $SO(5)$ symmetry that combines the AFM and VBS order parameters~\cite{Nahum2015PRL,lou2009antiferromagnetic,wang2022scaling,Jun2024SO(5)}. 

In this work, we use SSE simulations~\cite{sandvik1991quantum,sandvik1999stochastic,yan2019sweeping,yan2022global} to study surface properties. The multi-replica trick is employed to study the imaginary-time evolution of the entanglement Hamiltonian~\cite{yan2023unlocking,wu2023classical,li2024relevant,song2023different}. Details can be found in the supplemental materials (SM)~\footnote{See Supplemental Material at [URL] for details of numerical methods, finite-size scaling analysis, and additional data.}. The edge energy spectra and the ES can be obtained from the related imaginary-time correlations through the stochastic analytic continuation (SAC) method~\cite{sandvik2016constrained,shao2017nearly,shao2023progress,yan2021topological}. 

Periodic boundary conditions are applied along one lattice direction, while open boundary conditions are used along the other direction when studying real surfaces. We consider two different surface configurations: the square lattices with standard cut and the $45^\circ$ tilted cut, as shown in Fig.~\ref{fig:model}(a) and (c). 

To study the entanglement Hamiltonian (EH), we consider standard and $45^\circ$ tilted square lattices with periodic boundary conditions applied in both lattice directions. To verify the applicability of the Li-Haldane-Poilblanc conjecture by comparing the properties of the surface Hamiltonian with those of the entanglement, we consider the entanglement region as half of the system and ensure that the entanglement Hamiltonian and the surface Hamiltonian have the same boundary configurations, as shown in Fig.~\ref{fig:model}(b) and (d). In our simulations, the inverse temperature scales as $\beta=2L$ to capture the ground-state properties.

\textit{\color{blue} Surface behaviors of DQCP.---}
To investigate the surface behaviors of the $J$-$Q_3$ model on two different lattices as Fig.~\ref{fig:model}(a) and (c) show, three types of observables are derived, (i): surface spin-spin correlation $C_{\parallel}(L)$ between two surface spins $i$ and $j$ with the longest distance $|i-j|=L/2$, (ii): surface-bulk spin-spin correlation $C_{\perp}(L)$ between one surface spin $i$ and the other $j$ moving a distance $L/2$ perpendicular to the surface, and (iii): surface Binder cumulant $U_{2}(L)= \frac{5}{6}\left(3-\frac{\langle m_{s}^z(L)^{4}\rangle}{\langle m_{s}^z(L)^{2}\rangle^{2}}\right)$~\cite{binder1981critical,sandvik2010computational,wang2022bulk}, where the surface magnetization $m_{s}^z =\sum_{i \in {\rm surface}} \phi_i S_i^z$, $\phi_i = \pm 1$ depending on which sublattice spin $i$ belongs to.

\begin{figure*}
    \centering
    \includegraphics[width=\linewidth]{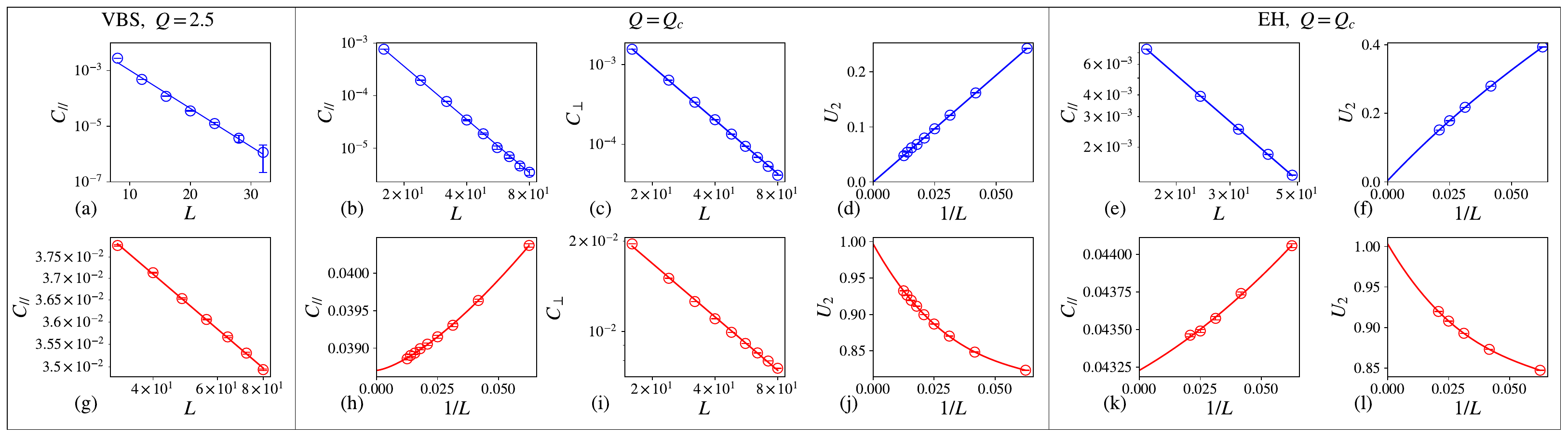}
    \caption{Finite-size scaling of observables for $J$-$Q_3$ model. Top row: standard (axis–aligned) cut (in blue); bottom row: $45^{\circ}$ tilted cut (in red). Columns (left to right) show (a)(g): surface correlations $C_{\parallel}$ at $Q = 2.5$ in VBS phase with open boundary condition (``real cut''), (b)(c)(d)(h)(i)(j): surface correlation $C_{\parallel}$, surface–bulk correlations $C_{\perp}$, and Binder cumulant $U_{2}$ at $Q = Q_c$ with real cut, (e)(f)(k)(l): $C_{\parallel}$ and $U_2$ of entanglement–Hamiltonian (``fake-cut'') at $Q = Q_c$, respectively. In the VBS phase, $C_{\parallel}$ decays to $0$ (a): exponentially and (g): algebraically. At the putative DQCP, in the standard geometry $C_{\parallel}$ and $U_{2}$ vanish algebraically, while the tilted edge saturates $C_{\parallel}\neq0$ and $U_{2}\to1$ when $L \to \infty$, signaling an extraordinary order that is mirrored in the entanglement Hamiltonian edge as well. The system sizes of the figures are (a): $L = 8 \sim 32$, (g): $L = 32 \sim 80$, (b)(c)(d)(h)(i)(j): $L = 16 \sim 80$, (e)(f)(k)(l): $L = 16 \sim 48$, respectively. Solid lines are least–squares fits described in the text.}
    \label{fig:order}
\end{figure*}

We first investigate the VBS phase of the $J$-$Q_3$ model at $Q = 2.5 > Q_c$. The surface correlation $C_{\parallel}(L)$ in the case with standard cut is plotted in Fig. \ref{fig:order}(a). The data can be fitted using straight lines on a linear-log scale, indicating that the correlations decay exponentially. By fitting the curves with $ C_{\parallel}(L) \sim \exp{(-L/a)}$, we obtain $a=3.1(2)$, which means that the surface state is gapped. Meanwhile, $C_{\parallel}(L)$ in the case with $45^\circ$ tilted cut at the same bulk phase exhibit completely different finite-size behavior, as shown in Fig.~\ref{fig:order}(g). We observe that $C_{\parallel}(L)$ decays with the system size $L$ algebraically, $C_{\parallel}(L) \sim  L^p$ with $p=-0.0850(6)$, indicating that the surface state is gapless while the bulk is a gapped VBS. 

According to surface‐criticality theory and extensive numerical studies at $(2+1)$-dimensional $O(N)$ bulk critical points, the boundary response depends on whether the surface itself is gapped or gapless when it reaches the bulk transition~\footnote{Although the boundary correlation function is always power law decay at bulk critical point which reflects the edge is gapless, here the ``gapped/gapless surface'' means the status of surface when the bulk is gapped but very close to the critical point.}. When the surface is initially gapped, its correlations at the bulk critical point are entirely inherited from the bulk and therefore decay algebraically with a larger exponent—i.e., they fall off faster—than the corresponding bulk correlations, a regime traditionally labeled “ordinary”~\cite{Cardy,deng2005surface,zhang2017}. If the surface hosts intrinsic gapless modes, two possibilities emerge: the gapless boundary may remain disordered yet flow with the bulk to a joint multicritical fixed point (the “special” transition), or develop spontaneous long-range order restricted to the boundary, producing an “extraordinary” transition in which ordered surface layers coexist with a critical bulk~\cite{ding2018,weber2018nonordinary,weber2019,wang2022bulk,wang2023extraordinary,zhu2021surface,wang2024unconventional,Jian2021,ding2023special}.

We find that the above rules still hold at the putative DQCP of $J$-$Q_3$ model. As shown in Fig.~\ref{fig:order}(b), the surface correlation function $C_{\parallel}(L)$ for the standard square lattice cut decays with system size $L$ according to a power law. We fit the data in the scaling form $C_{\parallel}(L) \sim L^{-(1+\eta_{\parallel})}$ yielding $\eta_{\parallel} = 2.36(2)$. The surface bulk correlation function $C_{\perp}(L)$ in Fig.\ref{fig:order}(c) also decays algebraically; fitting it to $C_{\perp}(L) \sim L^{-(1+\eta_{\perp})}$ we obtain $\eta_{\perp} = 1.305(10)$. We find that the two exponents obey the scaling law $2\eta_\perp = \eta_\parallel +\eta$, where $\eta = 0.26(3)$ calculated in previous work~\cite{sandvik2007evidence} is the anomalous magnetic scaling dimension at the bulk critical point~\cite{Diehl}. This result falls outside the known categories of surface critical exponents (see SM), indicating a new surface universality class.

In stark contrast, as shown in Fig.~\ref{fig:order}(h)(i), the surface correlation function $C_{\parallel}(L)$ for the $45^\circ$ tilted cut remains finite as $L \to \infty$, while the surface bulk correlation function $C_{\perp}(L)$ still decays algebraically to $0$, indicating that the gapless surface state merges with the bulk criticality, leading to the formation of a long-range order with symmetry breaking on the surface. Here we also try to extract the critical exponents and constant $b$ associated with symmetry breaking, using formulas $C_{\parallel}(L) = kL^{-(1+\eta_{\parallel})}+b$ and $C_{\perp}(L) \sim L^{-(1+\eta_{\perp})}$. We find $\eta_{\parallel} = 0.43(6)$ and $\eta_{\perp} = -0.461(1)$ and $b = 0.03870(3)$.

To further determine the long-range order on the surface, we calculate the surface Binder cumulant $U_{2}(L)$. If $U_{2}(L)$ converges to 1 as $L \to \infty$, it indicates the presence of a magnetic order.  Conversely, if $U_{2}(L)$ approaches $0$ with increasing system size, it implies that the system is in a disordered phase. The numerical results of $ U_{2}(L)$ for the $45^\circ$ tilted geometry as a function of size $1/L$ are plotted in Fig.~\ref{fig:order}(j). We fit the data using a polynomial of $1/L$: $U_{2}(L) = c_0+c_{1}L^{-1}+c_{2}L^{-2}+c_{3}L^{-3}$ and find a statistically sound estimation $c_0 = 0.9971(7)$. This further supports the existence of long-range order on the surface. To facilitate the comparison, we also calculated $ U_{2}(L)$ for the standard cut case, as shown in Fig.~\ref{fig:order}(d). We find $U_{2}(L) \to 0$ in the thermodynamic limit with a linear fitting $U_{2}(L) = b + kL^{-1}$, the intercept of which is $b = -0.0006(3)$. This is consistent with the conclusion from the correlation functions that the surface does not exhibit symmetry breaking in this case. To further clarify the scaling behavior of $C_{\parallel}(L)$ and the convergence of $U_2(L)$, additional finite-size analyses are presented in the SM.

The above data display that the tilted cut induces an extra gapless mode on the edge and further leads to an extraordinary surface criticality, while the standard cut does not. The next question is whether the extraordinary surface criticality will affect the entanglement behaviors even under an entanglement cut instead of a real cut.

\textit{\color{blue} Surface behaviors of the entanglement Hamiltonian.---}
We also compute the surface–surface correlations $C_{\parallel}(L)$ and the surface Binder cumulant $U_{2}(L)$ of the ground state of the entanglement Hamiltonian for both bipartitions via the multi-replica method~\cite{li2024relevant,wang2025sudden}, more technical details are in the SM. This directly confirms that the surface critical behavior on a real boundary is mirrored at the entanglement edge and thus impacts entanglement-based diagnostics~\footnote{Although the entanglement–Hamiltonian calculations are limited to $L_{max} = 48$, they nonetheless reproduce the same finite-size trends observed on the real boundary}. For Fig.~\ref{fig:order}(e), we perform finite-size extrapolations using the form $C_{\parallel}(L) = kL^{-(1+\eta_{\parallel})}$, and get $\eta_{\parallel} = 0.525(2)$. A second-order polynomial is used to fit Fig.~\ref{fig:order}(k) and yields a finite value of $C_{\parallel}(L)$ at the thermodynamic limit $L \to \infty$, which is $c_0 = 0.04322(5)$. Fig.~\ref{fig:order}(f)(l) are fitted both with a third-order polynomial whose intercepts are $c_0 = 0.004(2)$ for the standard cut, and $c_0 = 1.003(2)$ for the tilted cut.

At first sight, a $45^\circ$ tilted bipartition seems advantageous because it does not preferentially intersect any of the four columnar VBS patterns, whereas an axis–aligned (standard) cut can bias them along the edge, as the Ref.~\cite{d2024entanglement} said. Our data, however, reveal that this geometric ``fairness’’ changes the boundary universality class: at the putative bulk critical point, the tilted cut flows to an extraordinary boundary that orders, while the standard cut flows to an ordinary boundary that remains disordered (Fig.~\ref{fig:order}). In the language of the emergent SO(5) description, these two geometries impose distinct boundary conditions—one that spontaneously breaks SO(5) at the edge and one that does not—so any entanglement observable along the cut can mix SO(5)-symmetric bulk critical fluctuations with symmetry-breaking surface contributions. In other words, the extra gapless edge mode also contributes to the entanglement, as it does in the real cut.

\textit{\color{blue} Entanglement spectrum and edge energy spectrum.---}
This raises a concrete question: is the EE and ES along a given cut merely a cleaner proxy for bulk $SO(5)$ criticality, or is it partly \emph{imprinted} by the boundary fixed point selected by that geometry? A direct way to diagnose this is through the Li–Haldane–Poilblanc correspondence: if the entanglement Hamiltonian resembles the edge Hamiltonian, then the low-lying ES should reproduce the physical edge excitation spectrum for the \emph{same} boundary geometry. Verifying this correspondence allows us to identify the boundary universality class seen by the ES and EE, in turn, assess how much of the “tilted advantage’’ originates from surface physics rather than the bulk.

\begin{figure}
    \centering
    \includegraphics[width=1.0\linewidth]{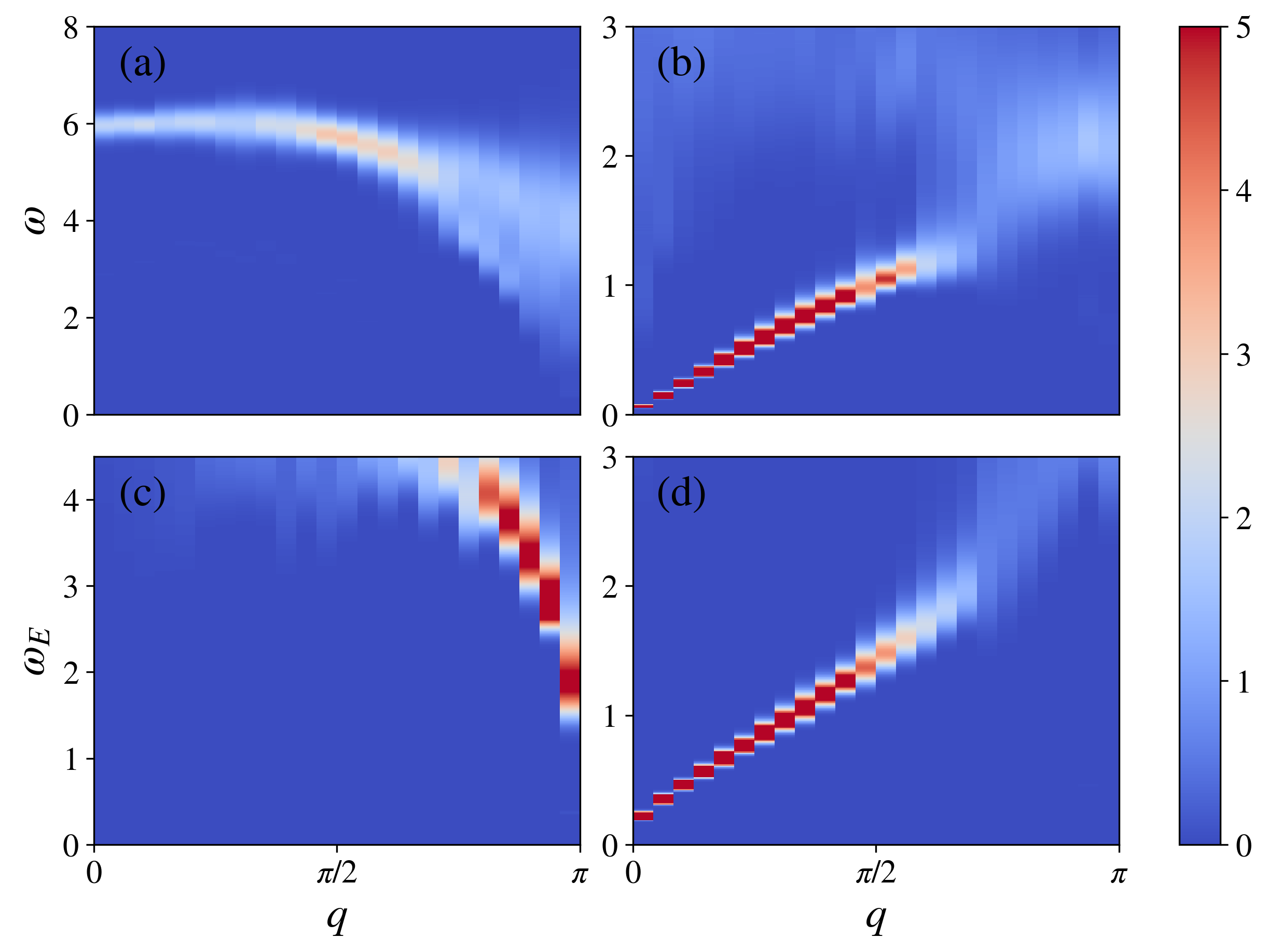}
    \caption{(a)(b) Edge excitation spectra with open boundary condition (real cut) and (c)(d) entanglement spectra (fake cut) of $J$-$Q_3$ model near the entanglement boundary at the putative DQCP. (a)(c) are using the standard cutting (bipartition), while (b)(d) are $45^\circ$ tilted. The system size $L = 48$ for all spectra. The system inverse temperature is $\beta = 96$ for excitation spectra and $\beta = 80$ for entanglement spectra. Note that (a)(b) are plotted using $\omega$ with the dimension of energy, while (c)(d) are plotted using a dimensionless variable $\omega_E$, resulting in the difference in vertical scales.}
    \label{fig:spectra}
\end{figure}

To test whether the ES resembles edge energy spectrum at a critical point, we examine both the surface excitation spectra (real cuts, see Fig.~\ref{fig:model} (a) and (c)) and the low-lying entanglement spectra (fake cuts, see Fig.~\ref{fig:model} (b) and (d)) of the $J$-$Q_3$ model. Fig.~\ref{fig:spectra}(a) and Fig.~\ref{fig:spectra}(b) present the dynamical structure factors on real surfaces of standard and tilted lattices, respectively. In both cases, we measured the imaginary time correlations at the entanglement/physical edge and obtain the related spectra via SAC. As discussed, for the standard cut, the criticality is entirely bulk-driven, so the boundary modes originate from the bulk critical modes. For the tilted cut, the coupling between gapless boundary modes in the bulk VBS phase and bulk critical modes induces a symmetry-breaking boundary state. Hence, the observed surface modes arise from the Goldstone mode due to spontaneous continuous symmetry breaking.

Fig.~\ref{fig:spectra}(c) and Fig.~\ref{fig:spectra}(d) show the corresponding entanglement spectra of both lattices. For both cases, the entanglement and surface energy spectra closely resemble each other, in qualitative agreement with the Li-Haldane-Poilblanc conjecture. We note that the entanglement spectrum for the tilted geometry is particularly sharp and well-resolved; for the standard case, while less clean due to broader spectral features, it still displays qualitatively similar dispersion to its physical surface counterpart. We argue that this is due to the correspondence between ES and edge energy spectrum holds when the bulk is gapped, but it would be corrected while the bulk becomes gapless~\cite{wang2025sudden,li2024relevant,liu2025worldline}. Our numerical results show that Li-Haldane-Poilblanc conjecture is still approximately, but not exactly, remaining at the critical point (it can be understood via a wormhole picture in the path integral of reduced density matrix, details are in the SM). It demonstrates that the edge effect is reflected in the entanglement.

These results reveal that ES in both standard and tilted geometries are affected by the physics of their surface critical behaviors. Crucially, in the tilted case, the boundary is ordered even at the critical point, and the ES necessarily contains contributions from both bulk criticality and surface ordering. Therefore, using EE, which serves as a summary of the entanglement spectrum, to characterize bulk criticality can be misleading. This observation casts doubt on prior interpretations of improved CFT matching via tilted bipartitions: what appears to be in better agreement with conformal scaling may instead stem from the inclusion of a distinct surface contribution.

\textit{\color{blue} Discussions and conclusions.---} 
We conclude that only the cut with ordinary edge faithfully captures the bulk critical properties; in contrast, cuts that generate nontrivial edge states yield entanglement contributions that mix bulk and boundary physics. In previous works~\cite{zhao2022scaling,zhao2022measuring,wang2025probing}, it was observed that the EE behaviors in the columnar dimerized Heisenberg model are consistent with field theory only when the entanglement bipartition boundary avoids all the dimers. The underlying reason, which was not fully clarified at the time, is now apparent: the entanglement cut avoiding all dimers they used realizes an ordinary boundary~\cite{ding2018}, which faithfully reflects the bulk criticality without introducing additional edge modes. Related numerical evidence also comes from disorder operator (a nonlocal measure similar to entanglement entropy) studies~\cite{liu2024measuring}, where boundary contributions at a special surface critical point simultaneously encode the bulk (2+1) D $O(3)$ universality and a boundary Luttinger liquid.

Finally, we emphasize that the practice of preserving the symmetry of the VBS order parameter in defining the bipartition~\cite{d2024entanglement} is reasonable and physically motivated. However, when such symmetry-preserving cuts necessarily introduce boundary order or gapless edge excitations, the resulting entanglement scaling must incorporate these boundary contributions~\cite{ma2022edge} rather than being attributed solely to the bulk.

In summary, our work has for the first time established a bridge between the two fields: surface criticality and many-body entanglement. Through numerical calculations, it was shown that the correspondence between the boundary energy spectrum and the entanglement spectrum still holds approximately even when the bulk is gapless, further indicating that the entanglement behavior is affected by the extra gapless edge. If one wants to ensure that entanglement entropy accurately reflects the critical behavior of the system, an ordinary cut corresponding to ordinary surface criticality is required. Otherwise, the extra gapless edge mode will also contribute to the entanglement behaviors, that is, the entanglement behaviors do not faithfully describe the bulk. The data supporting the findings of this article are openly available at~\cite{zhu2025zenodo}.

\begin{acknowledgements}
\textit{\color{blue} Acknowledgments.---} 
YCW and ZY acknowledge the collaboration with Jiarui Zhao, Meng Cheng and Zi Yang Meng in another related work. 
YCW acknowledges the support from the Natural Science Foundation of China (Grant No. 12474216) and Zhejiang Provincial Natural Science Foundation of China (Grant No. LZ23A040003), and the Start-up Funding of Hangzhou International Innovation Institute of Beihang University. 
ZW is supported by the China Postdoctoral Science Foundation under Grants No.2024M752898. ZL is supported by the China Postdoctoral Science Foundation under Grants No.2024M762935 and NSFC Special Fund for Theoretical Physics under Grants No.12447119. 
The work is supported by the Scientific Research Project (No.WU2025B011) and the Start-up Funding of Westlake University.
The authors thank the high-performance computing centers of Westlake University and of Hangzhou International Innovation Institute of Beihang University and the Beijng PARATERA Tech Co.,Ltd. for providing HPC resources.
\end{acknowledgements}

\bibliography{jqcut}

\end{document}

\title{Bipartite entanglement and surface criticality:\\The extra contribution of non-ordinary edge in entanglement\\-Supplementary Material-}

\author{Yanzhang Zhu}
\affiliation{State Key Laboratory of Surface Physics and Department of Physics, Fudan University, Shanghai 200433, China}
\affiliation{Department of Physics, School of Science and Research Center for Industries of the Future, Westlake University, Hangzhou 310030, China}
\affiliation{Institute of Natural Sciences, Westlake Institute for Advanced Study, Hangzhou 310024, China}

\author{Zenan Liu}
\affiliation{Department of Physics, School of Science and Research Center for Industries of the Future, Westlake University, Hangzhou 310030, China}
\affiliation{Institute of Natural Sciences, Westlake Institute for Advanced Study, Hangzhou 310024, China}

\author{Zhe Wang}
\email{wangzhe90@westlake.edu.cn}
\affiliation{Hangzhou International Innovation Institute, Beihang University, Hangzhou 311115, China}
\affiliation{Department of Physics, School of Science and Research Center for Industries of the Future, Westlake University, Hangzhou 310030, China}
\affiliation{Institute of Natural Sciences, Westlake Institute for Advanced Study, Hangzhou 310024, China}

\author{Yan-Cheng Wang}
\email{ycwangphys@buaa.edu.cn}
\affiliation{Hangzhou International Innovation Institute, Beihang University, Hangzhou 311115, China}
\affiliation{Tianmushan Laboratory, Hangzhou 311115, China}

\author{Zheng Yan}
\email{zhengyan@westlake.edu.cn}
\affiliation{Department of Physics, School of Science and Research Center for Industries of the Future, Westlake University, Hangzhou 310030, China}
\affiliation{Institute of Natural Sciences, Westlake Institute for Advanced Study, Hangzhou 310024, China}

\date{\today}
\maketitle

In this Supplementary Material, we present detailed descriptions of the  stochastic series expansion (SSE) Quantum Monte Carlo (QMC) implementation of the entanglement Hamiltonian (EH) using the multi-replica trick, the extraction of the entanglement spectrum (ES) through stochastic analytic continuation (SAC), the theoretical framework linking the EH to the physical edge via the wormhole picture, the supplementary boundary analysis supporting our main results, and the comparison of surface critical exponents across different universality classes.

\section{Simulation of the Entanglement Hamiltonian}

\subsection{Replica construction}

\begin{figure}
    \centering
    \includegraphics[width=0.8\linewidth]{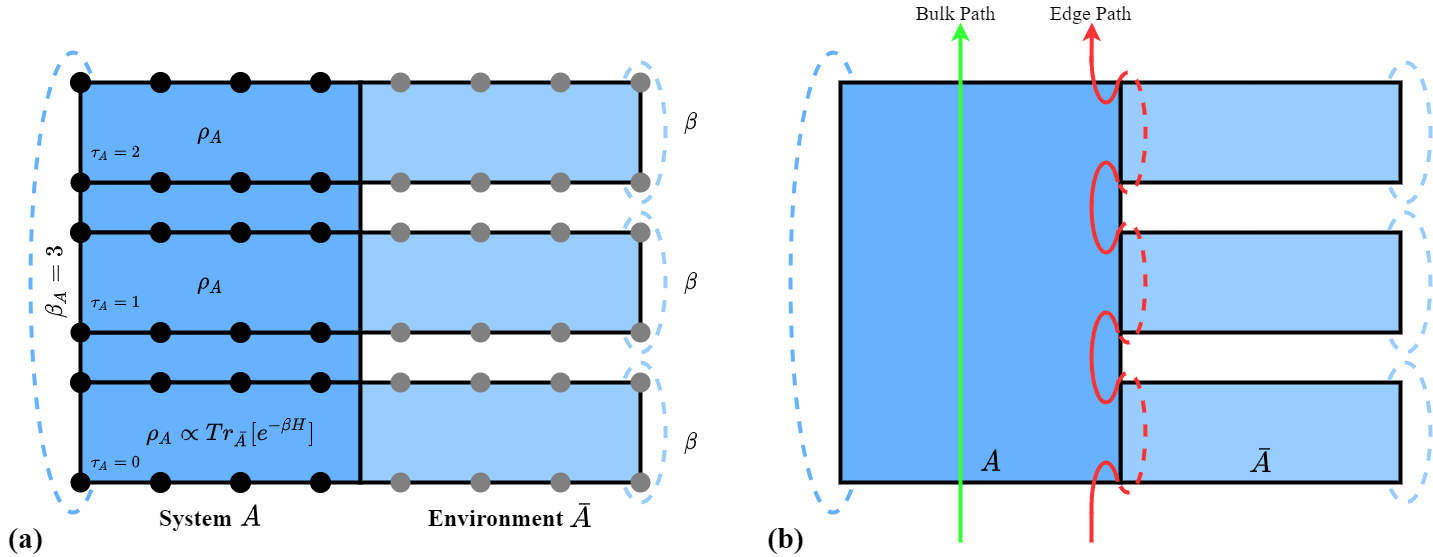}
    \caption{Illustration of the replica manifold and the origin of the wormhole picture in the path integral of reduced density matrix. (a) Three replicas are shown for the manifold of path integral, each of imaginary–time extent (inverse temperature) $\beta$. Blue dashed lines indicate periodic boundary condition of imaginary time. The environment $\bar{A}$ (light blue) is traced out, so within each replica its imaginary-time boundary is closed (top connected to bottom). (b) Schematic diagram of worldline paths. The bulk path (green) crosses the entire replica, costing an imaginary-time length proportional to $\beta$ (inverse temperature of the total system, the system approaches to the ground state when $\beta\rightarrow \infty$), whereas the edge path (red) connects the lower and upper boundaries of $\bar{A}$ directly, costing only a very short length. The edge path actually provides a lower-cost channel in the path integral, thereby the paths around the entanglement edge take larger weights in the path integral.}
    \label{fig:eh}
\end{figure}

We access the EH of a spatial subregion $A$ by sampling the replicated partition function
\begin{equation}
  Z_n(A)\propto\mathrm{Tr}\,\rho_A^n=\mathrm{Tr}\!\left[e^{-nH_{A}}\right],
\end{equation}
Here, the reduced density matrix $\rho_A=e^{H_{A}}$ and $H_{A}$ is entanglement Hamiltonian. This partition function can be implemented as an $n$-replica path integral in which the worldlines in $A$ are glued cyclically between replicas, while those in the environment $\bar A$ are traced independently on each replica. This is the key identification underlying the wormhole picture: observables built from operators supported entirely in $A$ and evaluated in the replicated ensemble are thermal expectation values of the EH at $\beta_A=n$. In the stochastic-series-expansion (SSE) representation this corresponds to imposing (i) a cyclic boundary condition in imaginary time $\tau$ across replicas for all sites in $A$ and (ii) the usual periodic boundary condition in $\tau$ on each replica for $\bar A$. Fig.~\ref{fig:eh} (a) illustrates the geometry for $n=3$: blue links indicate the inter-replica identifications in $A$, gray regions denote $\bar A$ (traced), and the black region denotes $A$ (glued). With this construction, operators supported in $A$ probe thermal correlation functions of the EH, i.e., the replica index $n$ plays the role of an \emph{effective} inverse temperature $\beta_A$ for $H_{A}$, providing a direct route to low-lying entanglement physics by increasing $n$~[48, 49, 57].

\subsection{SSE implementation and observables}

We simulate each replica with the standard SSE operator-string formalism with diagonal and loop updates. For sites in $A$, the head/tail worldline connectivity at $\tau=\beta$ on replica $k$ is joined to $\tau=0$ on replica $k{+}1$ (mod $n$); for $\bar A$, $\tau\!=\!\beta$ connects back to $\tau\!=\!0$ on the same replica. Space is periodic along the cut direction. Unless stated otherwise, we use $\beta=L$ to project the ground state of $H$ on each replica, and a large replica number $n$ (e.g., $n = \beta_A \sim 2L$) for EH to resolve the lowest entanglement modes. All measurements are accumulated after equilibration across $50000$ SSE sweeps, with multiple seeds for error estimates.

Within each replica ensemble we measure equal-time correlators of surface spins and the binder cumulant supported on the entanglement bipartition edge $\partial A$:
\begin{align}
  &\text{Surface–surface:}\quad 
  C_{\parallel}(L) \equiv \big\langle S^z_i \cdot S^z_j \big\rangle, \quad i,j\in \partial A,\ |i{-}j|\!=\!L/2,\\
  &\text{Binder cumulant:}\quad 
  U_{2}(L)= \frac{5}{6}\left(3-\frac{\langle m_{s}^z(L)^{4}\rangle}{\langle m_{s}^z(L)^{2}\rangle^{2}}\right),\ m_{s}^z =\sum_{i \in {\rm \partial A}} \phi_i S_i^z.
\end{align}
To obtain the entanglement spectrum (ES) with stochastic analytic continuation (SAC), we accumulate the imaginary time spin correlations along the entanglement edge,
\begin{equation}
  G_{q}(\tau)=\Big\langle S^z_{-q}(\tau)\,S^z_q(0)\Big\rangle,
  \qquad
  S^z_q(\tau) \equiv \frac{1}{\sqrt{N}}\sum_{i\in \partial A} e^{\,\mathrm{i}q\cdot r_i}\,S^z_i(\tau),
\end{equation}
where $\partial A$ denotes the set of boundary sites of the subsystem, $N=L_x$ is the number of spins on the bipartition surface $\partial A$ (suppose the entanglement bipartition is along the x axis), and $q=2\pi k/L_x$ ($k=0,\dots,L_x-1$). It is worth mentioning that, here we chose to do the Fourier transform first in the simulation, and then calculate the imaginary time correlation of spins in momentum space. It is also fine to record all data of imaginary time correlation in real space during the simulation, and carry out the transform into $k$ space afterwords for SAC. 

\subsection{Stochastic analytic continuation}
Here we only briefly discuss how SAC is carried out to obtain the entanglement spectrum. For those who are interested in actually resolving some energy spectra with QMC method, we highly recommend reading Ref.~[61, 63]. 
Real–frequency spectra are obtained from QMC imaginary–time data by inverting the integral equation
\begin{equation}
G(\tau)=\int_{0}^{\infty}\!d\omega\,A(\omega)\,K(\tau,\omega),\qquad 
K(\tau,\omega)=\frac{e^{-\tau\omega}+e^{-(\beta-\tau)\omega}}{\pi\,[1+e^{-\beta\omega}]},
\end{equation}
with $A(\omega)\!\ge\!0$ the bosonic spectral function (e.g., the dynamical spin structure factor) and $G(\tau)=\langle O^{\dagger}(\tau)O(0)\rangle$ measured for $\tau\in[0,\beta]$ satisfying $G(\tau)=G(\beta-\tau)$. We parametrize $A(\omega)$ as a sum of $\delta$–functions on a dense grid and sample it stochastically by minimizing the correlated misfit
\begin{equation}
\chi^{2}[A]=\sum_{i,j}\big(G_{i}-\overline{G}_{i}\big)\,C^{-1}_{ij}\,\big(G_{j}-\overline{G}_{j}\big),
\end{equation}
where $\overline{G}_{i}$ are QMC means at times $\tau_{i}$ and $C$ is their covariance (estimated from binning/bootstrapping). Trial spectra are drawn with weight
\begin{equation}
P[A]\propto\exp\!\left[-\frac{\chi^{2}[A]}{2\,\Theta}\right],
\end{equation}
with the sampling “temperature’’ $\Theta$ tuned so that $\langle\chi^{2}\rangle/N_{\tau}\!\approx\!1$. Positivity and optional constraints (sum rules, support windows, or edge shapes) stabilize the inversion; the final spectrum $\langle A(\omega)\rangle$ and its uncertainty are obtained by averaging over the posterior ensemble. In practice, we work with $\tau\in[0,\beta/2]$ and use the kernel above so that only $\omega\!\ge\!0$ enters the continuation. Therefore, if we get the imaginary-time correlation function for entanglement Hamiltonian, we can use SAC to capture the entanglement spectrum via importing the corresponding correlation function data.

\section{The wormhole picture as a potential explanation of Li-Haldane-Poilblanc correspondence}

The Li–Haldane–Poilblanc conjecture proposes that the low-lying entanglement spectrum (ES) of a quantum many-body system resembles the energy spectrum of its physical edge states in a gapped system, establishing a correspondence between the entanglement Hamiltonian and the boundary Hamiltonian of the original system. A recently proposed \textbf{wormhole picture} in the path integral of reduced density matrix~[57] provides an intuitive geometrical understanding of this correspondence. 

As depicted in Fig.~\ref{fig:eh}(b) here, each replica represents a reduced density matrix $\rho_A\propto \mathrm{Tr}_{\bar{A}}[e^{-\beta H}]$. Using the replica trick, we construct the generalized partition function $\mathcal{Z}^{(n)}_{A}\propto \mathrm{Tr}\{\rho^n_A\}$ within the path integral framework. The trace of environment $\bar{A}$ connects the upper and lower imaginary-time boundaries of each replica. It provides a convenient way for worldlines near entanglement edge to get to the upper edge from the lower edge with little cost, dubbed wormhole picture. As a result, worldlines near the entangled edge avoid traversing the whole replica with huge cost (the imaginary time length in the replica $\sim \beta$, $\beta\rightarrow\infty$ since we are discussing the ground state).  

In the path-integral language, the cost of each path is proportional to $\overline{\mathcal{L}}\times\Delta(\overline{\mathcal{L}})$, where $\overline{\mathcal{L}}$ is the average imaginary-time path length and $\Delta(\overline{\mathcal{L}})$ denotes the average excitation gap along that path. A lower cost leads to a larger path weight because the weight is $e^{-\overline{\mathcal{L}}\times\Delta(\overline{\mathcal{L}})}$, which consequently makes a greater contribution to the low-lying entanglement spectrum (ES). Since the imaginary time-length in one replica is equal to $\beta$, while the effective time-length around the entangled edge reduces to a constant of $\mathrm{O}(1)$ (we set it as $1$) due to the wormhole effect, thus the ratio between the bulk path's and edge path's time-length can be estimated as $\beta:1$.

By considering both the path length and the excitation gaps, the relative cost between bulk and edge paths can be estimated as $\beta\Delta_b:\Delta_e$, where $\Delta_b$ and $\Delta_e$ are the bulk and edge gaps, respectively. In the ground-state limit, $\beta\rightarrow\infty$ which renders $\beta\Delta_b \gg \Delta_e$ if both gaps are finite. This indicates that the edge paths (red lines in Fig.~\ref{fig:eh}(b)) has lower cost than the imaginary-time bulk paths (green lines in Fig.~\ref{fig:eh}(b)), which has a larger weight and contributes more to the low-lying ES. Therefore, the edge excitation mode mainly contributes to the low-lying ES, which provides a natural and quantitative explanation of the Li–Haldane–Poilblanc conjecture.

The wormhole picture can be extended to a variety of magnetic systems without topology, including the spin-1/2 Heisenberg ladder and the bilayer Heisenberg model~[47, 57]. Even when the boundary is gapped and the conventional bulk–edge correspondence is broken, the wormhole picture remains valid~[48]. In this case, as $\beta\rightarrow\infty$ and gaps are finite, the inequality $\beta\Delta_b>\Delta_e$ still holds, so the entanglement spectrum continues to resemble the edge-state spectrum of the gapped boundary. This demonstrates that the entanglement Hamiltonian retains correspondence with the boundary Hamiltonian of the original system.

For gapless systems of finite size systems, actually the finite gaps scale with $L^{-z}$ where $z$ is the dynamical exponent. Since we set $\beta=L$ ($z=1$ here) in our work, thus the finite gap $\sim 1/\beta$. Now the order of the ratio  $\beta\Delta_b:\Delta_e$ roughly becomes $1: 1/\beta$, which still guarantees that the edge contributes more in the low-lying ES.
As a result, the low-lying ES is still similar to the physical edge spectrum, preserving the validity of the wormhole picture at criticality. Similar behavior has been observed in the two-dimensional Affleck–Kennedy–Lieb–Tasaki model~[74]. This framework thus explains why, at the deconfined quantum critical point in the $J$–$Q_3$ model, the correspondence between the entanglement spectrum and the physical edge spectrum persists even when the bulk is critical.

\section{Supplementary Boundary Analysis}

To clarify the scaling behaviors of boundary correlations and to answer potential doubts about the plotting scale used in main text, we present additional numerical analyses of the surface correlation function $C_{\parallel}(L)$ and the Binder cumulant $U_2(L)$ for the tilted geometry of the boundary. These data further confirm the conclusions summarized in the main text.

\begin{figure}[t]
\centering
\includegraphics[width=0.8\linewidth]{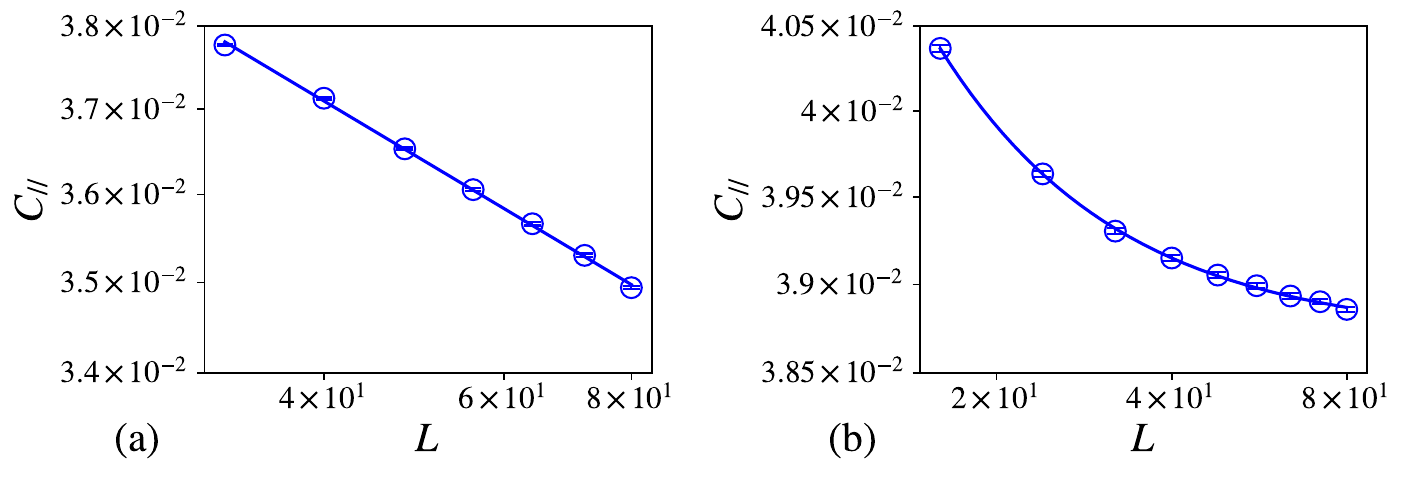}
\caption{(a) Surface correlation function $C_{\parallel}(L)$ in the VBS phase ($Q=2.5$), same as Fig.~2(g) of the main text. The data are well fitted by $C_{\parallel}(L) \sim  L^{-0.0850(6)}$. (b) Surface correlation function $C_{\parallel}(L)$ at the critical point ($Q=Q_c=1.49153$), using the same data as Fig.~2(h) of the main text but replotted here on a log-log scale for direct comparison with (a). The fit gives $C_{\parallel}(L)=0.08(2)L^{-1.43(6)}+0.03870(3)$, highlighting that, unlike the VBS case, the data approach a finite value as $L \to \infty$.}
\label{fig:loglog}
\end{figure}

Figure~\ref{fig:loglog} provides a direct log-log comparison of the surface correlation function $C_{\parallel}(L)$ in the VBS phase and at the critical point. In panel (a), corresponding to $Q=2.5$, the data follow a clean power law $C_{\parallel}(L)\sim L^{-0.0850(6)}$, which decays toward zero as $L$ increases. This behavior indicates that the boundary in the VBS phase is disordered but close to critical, with a very small decay exponent leading to an apparent plateau on limited system sizes. Such a slow decay can easily be mistaken for a finite saturation when plotted against $1/L$, hence the log-log presentation clarifies that the correlation ultimately vanishes in the thermodynamic limit. 

Panel (b) shows the same observable at the critical coupling $Q_c=1.49153$, plotted on the same scale for direct comparison. Here the data are fitted by $C_{\parallel}(L)=0.08(2)L^{-1.43(6)}+0.03870(3)$, where the constant term dominates at large $L$. This clear convergence to a finite value as $L\to\infty$ signals that the boundary remains ordered even when the bulk is critical. The contrast between panels (a) and (b) highlights the essential physics of the tilted geometry: while the tilted boundary in the VBS phase is disordered and follows a weak power law, the tilted boundary develops its own long-range order at the DQCP, realizing an extraordinary surface critical state.

\begin{figure}[t]
\centering
\includegraphics[width=0.4\linewidth]{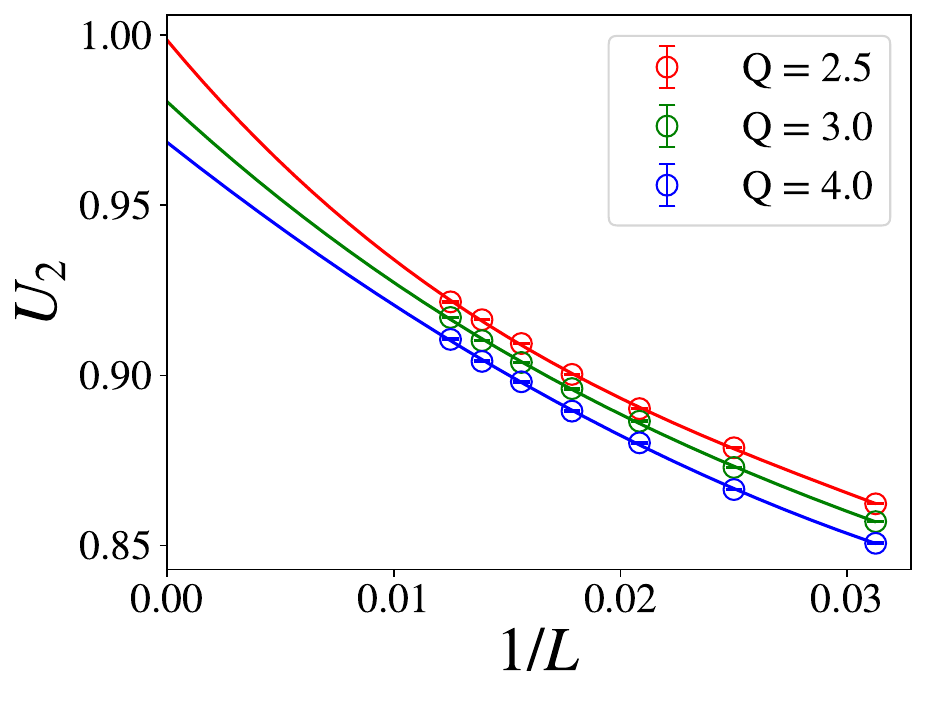}
\caption{Surface Binder cumulant $U_2(L)$ for the tilted boundary in the VBS phase at $Q=2.5$, $3.0$, and $4.0$. Solid lines denote third-order polynomial fits in $1/L$, yielding $U_2(\infty)=0.997(4)$, $0.978(5)$, and $0.968(7)$ respectively. As $Q$ increases (deeper into the VBS phase), $U_2$ decreases slightly from unity but remains finite, indicating a stable yet non-saturated boundary order.}
\label{fig:binder}
\end{figure}

Other complementary evidence is provided by the Binder cumulant results in Fig.~\ref{fig:binder}.  
Three representative couplings within the VBS phase, $Q=2.5$, $3.0$, and $4.0$, are analyzed.  
For each, $U_2(L)$ is fitted by a cubic polynomial in $1/L$, and the intercept at $1/L \to 0$ is extracted. The extrapolated values $U_2(\infty)=0.997(4)$, $0.978(5)$, and $0.968(7)$ fall well within the physically expected range between 0 and 1. This confirms that the Binder cumulant approaches a non-trivial, finite constant in the thermodynamic limit.  

As the system moves deeper into the VBS phase (larger $Q$), $U_2(\infty)$ exhibits a mild downward trend, signifying a gradual weakening of the boundary magnetization fluctuations but not their disappearance. The smooth evolution of $U_2$ with $Q$ is consistent with the picture that the tilted boundary remains ordered but its fluctuation amplitude depends on the relative strength of dimerization and interlayer coupling. Taken together with the correlation results in Fig.~\ref{fig:loglog}, these analyses confirm that the entanglement boundary exhibits long-range order both inside the VBS phase and at the DQCP, while its quantitative signatures evolve continuously across the transition. Such behavior strengthens the interpretation that the boundary degrees of freedom follow the universality of the extraordinary surface transition.

\section{Surface critical exponents}

Current studies of surface critical behavior (SCB) have primarily focused on three-dimensional O(n) critical points, especially for n = 1 (Ising model), n = 2 (XY model), and n = 3 (Heisenberg model). To the best of our knowledge, investigations into cases with n > 3 remain scarce. In Table~\ref{exp3}, we summarize the commonly recognized surface universality classes for reference and comparison.

\begin{ruledtabular}
  \begin{table*}[!h]
  \caption{ 
  For the reader's convenience, the surface critical exponents obtained in this work, as well as those at critical points in the 3D O(n) [n = 1, 2, and 3] universality class, are listed for comparison. Spe. indicates the special surface critical behavior. Ord. indicates the ordinary surface critical behavior.}
  \label{exp3}
  \begin{tabular}{c c c } 
SCB class at 3D O(3) critical point~[37-42, 44, 45]                               &$\eta_\parallel$       &  $\eta_\perp$    \\
  \hline
Spe.                                &   -0.56(1)      &  -0.259(3)   \\
Ord.                              &   1.36(6)     &  0.69(3)   \\
\hline
 SCB class at 3D O(2) critical point~[41, 69]                               &$\eta_\parallel$       &  $\eta_\perp$    \\
  \hline
Spe.                                &   -0.35(1)      &     \\
Ord.                              &   1.43(1)      &  0.69(2)   \\
\hline

\hline
 SCB class at 3D Ising critical point~[43, 69]                               &$\eta_\parallel$       &  $\eta_\perp$    \\
  \hline
Spe.                                &   -0.27(1)      &     \\
Ord.                              &   1.52(1)      &  0.82(8)   \\
\hline
 SCB class at DQCP (obtained in this work)                               &$\eta_\parallel$       &  $\eta_\perp$    \\
  \hline
Ord.                              &   2.36(2)      &  1.305(10)   \\
  \end{tabular}
  \end{table*}
\end{ruledtabular}